\documentclass[10pt, twocolumn]{wlscirep}
\usepackage[utf8]{inputenc}
\usepackage[T1]{fontenc}
\title{Diffusion-Based Electrocardiography Noise Quantification via Anomaly Detection}

\author[1,2]{Tae-Seong Han}
\author[1,2]{Jae-Wook Heo}
\author[1]{Hakseung Kim}
\author[1,2]{Cheol-Hui Lee}
\author[4]{Hyub Huh}
\author[5]{Eue-Keun Choi}
\author[6,7]{Hye Jin Kim}
\author[1,2,3,*]{Dong-Joo Kim}

\affil[1]{Department of Brain and Cognitive Engineering, Korea University, Seoul 02841, Republic of Korea}
\affil[2]{Interdisciplinary Program in Precision Public Health, Korea University, Seoul 02841, Republic of Korea}
\affil[3]{Department of Neurology, Korea University College of Medicine, Seoul 02841, Republic of Korea}
\affil[4]{Department of Anesthesiology and Pain Medicine, Kyung Hee University Hospital at Gangdong, Kyung Hee University College of Medicine, Seoul 05278, Republic of Korea}
\affil[5]{Department of Internal Medicine, Seoul National University College of Medicine and Seoul National University Hospital, Seoul 03080, Republic of Korea}
\affil[6]{Department of Biomedical Engineering, Yonsei University, Wonju 26493, Republic of Korea}
\affil[7]{Center for Nanoparticle Research, Institute for Basic Science, Seoul 08826, Republic of Korea}
\affil[*]{dongjookim@korea.ac.kr}

\begin{abstract}
Electrocardiography (ECG) signals are frequently degraded by noise, limiting their clinical reliability in both conventional and wearable settings. Existing methods for addressing ECG noise, relying on artifact classification or denoising, are constrained by annotation inconsistencies and poor generalizability. Here, we address these limitations by reframing ECG noise quantification as an anomaly detection task. We propose a diffusion-based framework trained to model the normative distribution of clean ECG signals, identifying deviations as noise without requiring explicit artifact labels. To robustly evaluate performance and mitigate label inconsistencies, we introduce a distribution-based metric using the Wasserstein-1 distance ($W_1$). Our model achieved a macro-average $W_1$ score of 1.308, outperforming the next-best method by over 48\%. External validation confirmed strong generalizability, facilitating the exclusion of noisy segments to improve diagnostic accuracy and support timely clinical intervention. This approach enhances real-time ECG monitoring and broadens ECG applicability in digital health technologies.
\end{abstract}

\begin{document}

\flushbottom
\maketitle
\thispagestyle{empty}

\section*{Introduction}

% figure-1
\begin{figure*}[!ht]
\centering
\includegraphics[width=0.90\linewidth]{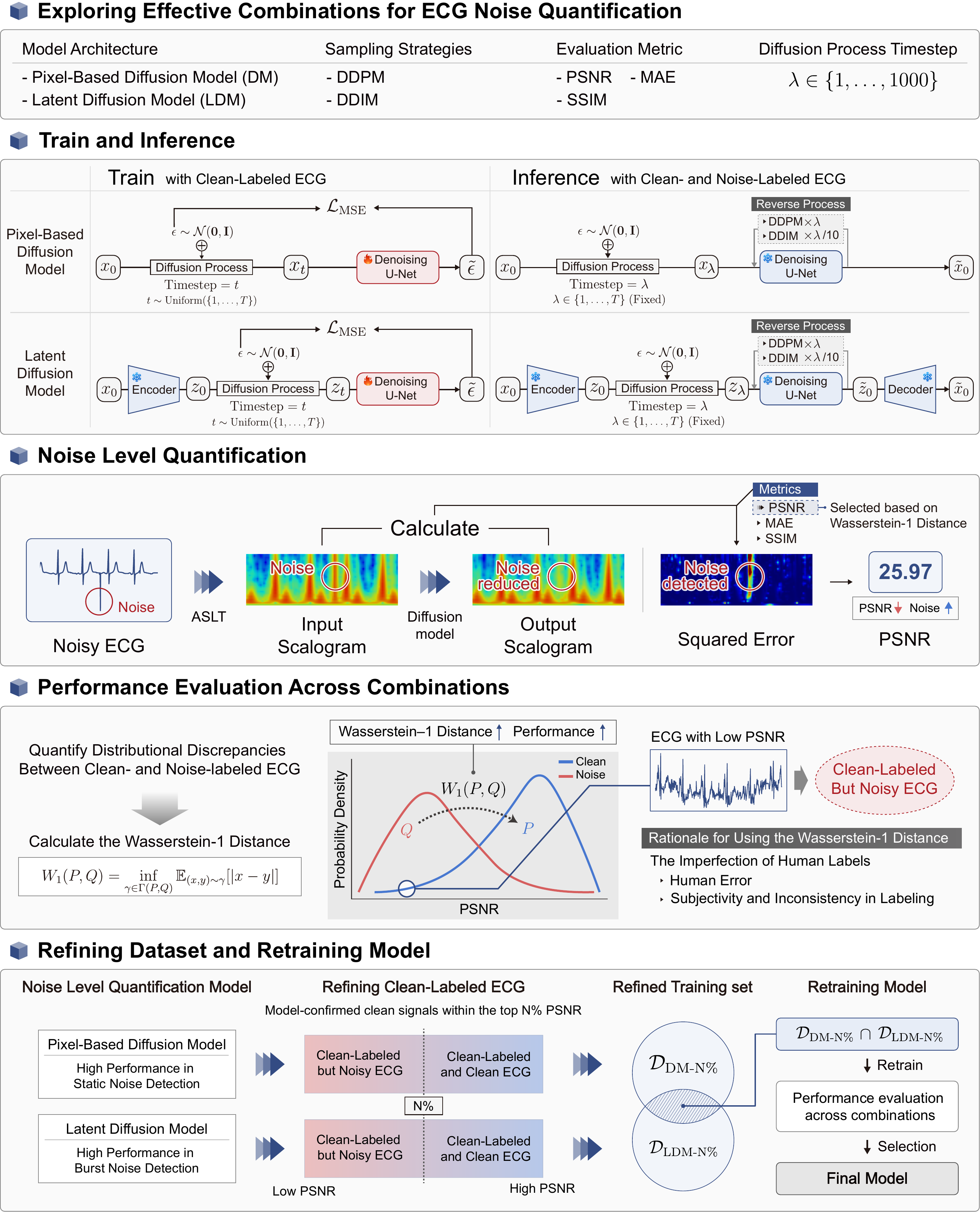}
\caption{
Overview of the proposed framework for ECG noise quantification using diffusion models. The framework comprises two architectures: a pixel-based diffusion model (DM) and a latent diffusion model (LDM), both trained to reconstruct clean ECG scalograms generated via the adaptive superlet transform (ASLT). During inference, sampling is performed using either denoising diffusion probabilistic models (DDPM) or denoising diffusion implicit models (DDIM). Noise quantification is assessed using image-based metrics, peak signal-to-noise ratio (PSNR), mean absolute error (MAE), and structural similarity index (SSIM), and measured via the Wasserstein-1 distance (\(W_1\)) between clean- and noise-labeled ECGs. A systematic comparison across model architectures, sampling strategies, diffusion steps (\(\lambda\)), and evaluation metrics guided optimal configuration selection and enabled iterative refinement through retraining on low-error samples.
}
\label{fig:figure1}
\end{figure*}

Electrocardiography (ECG) is a non-invasive technique essential for diagnosing and monitoring cardiac conditions. The proliferation of wearable devices has led to an increase in real-time ECG data collection, enabling continuous monitoring in both clinical and ambulatory environments~\cite{ref13,ref37,ref38}. However, ECG signals are highly susceptible to noise caused by external factors such as patient motion, electromyographic activity, and electrical interference, which can significantly degrade signal fidelity and complicate clinical interpretation~\cite{ref36}. Poor-quality signals not only compromise diagnostic accuracy and may trigger false alarms in critical care settings~\cite{ref39}. Ensuring the quality of such large-scale data remains a significant challenge, as manual inspection is impractical at this scale. Consequently, automated methods for ECG noise quantification are crucial for ensuring clinical reliability.

Despite its importance, the advancement of automated ECG noise quantification has been limited by several challenges. One major hurdle is the lack of large-scale, high-quality datasets specifically labeled for noise, which hampers robust model training. In addition, the inherent variability in ECG signals arising from physiological differences, electrode placement, and device-specific factors makes it difficult to develop models that generalize well across diverse conditions~\cite{ref1}. Furthermore, ECG labeling is prone to label noise owing to subjective interpretation, limited domain expertise, and the high cost of expert labeling~\cite{ref15,ref40,ref62}. 

Although various approaches have been proposed to address ECG noise, the current methodologies exhibit notable limitations. Traditional denoising methods such as Kalman filtering~\cite{ref2}, wavelet transforms~\cite{ref3}, and empirical mode decomposition~\cite{ref4} rely on fixed assumptions about the signal or noise characteristics. More recently, deep learning-based denoising methods have been trained to learn noise removal by adding synthetic noise to clean ECG signals~\cite{ref5,ref6,ref7}. However, these methods assume the presence of clean reference signals, which limits their performance when the ECGs are severely corrupted or missing, resulting in unrealistic reconstructions. Classification-based approaches, including rule-based algorithms~\cite{ref8,ref9}, signal quality indices (SQIs)~\cite{ref11,ref12, ref26, ref28}, and deep feature extractors~\cite{ref41,ref42}, are also constrained by annotation inconsistencies and label noise. These challenges highlight the need for a methodological shift. Here, we reframe ECG noise quantification as an anomaly detection problem. Instead of training a model to recognize a finite set of predefined noise categories, we train it to capture a comprehensive representation of clean ECG morphology. Any signal that deviates from this learned normative distribution is consequently identified as noise without requiring explicit artifact labels. This approach is intrinsically more robust to the unmodeled and unpredictable artifacts encountered in real-world clinical and ambulatory settings.

Among the various anomaly detection strategies~\cite{ref52,ref53,ref54,ref55}, reconstruction-based approaches have garnered considerable interest owing to their interpretability, which allows for direct comparison between reconstructed and original samples while also modeling in-distribution data without requiring explicit labels or prior knowledge of anomalies~\cite{ref16,ref17,ref18}. In these methods, a model is trained to reconstruct in-distribution samples, flagging inputs that yield high construction errors during inference as potential anomalies. However, these methods have limitations related to the information bottleneck. If the bottleneck is too narrow, in-distribution samples cannot be properly reconstructed; if it is too wide, out-of-distribution inputs may also be reconstructed with a low error, thereby blurring the distinction between normal and abnormal data~\cite{ref61}.

Diffusion models have emerged as promising alternatives to traditional reconstruction-based approaches for anomaly detection~\cite{ref19,ref20,ref21,ref22}. These models learn to reverse a gradual noising process, effectively denoising the inputs to match the distribution of clean data. Unlike fixed-bottleneck architectures, diffusion models offer adaptive reconstruction. By controlling the degree of corruption in the input, they can either generate new samples from pure noise or recover meaningful structures from partially corrupted inputs. This flexibility helps distinguish normal from anomalous data without relying on rigid encoding constraints~\cite{ref19}.

Building on this foundation, this study introduces a diffusion model-based anomaly detection framework for ECG noise quantification and automated data quality assessment to address the annotation inconsistencies and limited generalizability of conventional methods. As illustrated in Figure~\ref{fig:figure1}, we first train a model on ECG signals labeled as clean and use it to identify and exclude mislabeled signals that are noisy but annotated as clean, which may have resulted from labeling inconsistencies or human error. Retraining on the refined dataset composed of reliably clean signals yielded a more accurate noise quantification model. We evaluated this final model across several external ECG datasets, demonstrating its robustness under diverse recording conditions. This approach offers a scalable and objective solution for ECG noise assessment, ensuring high-quality ECG data for clinical and wearable device applications.

%%%%%%%%%%%%%%%%%%%%%%%%%%%%%%%%%%%%%%%%%%%%%%%%%%%%%%%%%%%%%%%%%%%%%%%

\section*{Methods}
\subsection*{Preprocessing}
\begin{figure}[!ht]
\centering
\includegraphics[width=\linewidth]{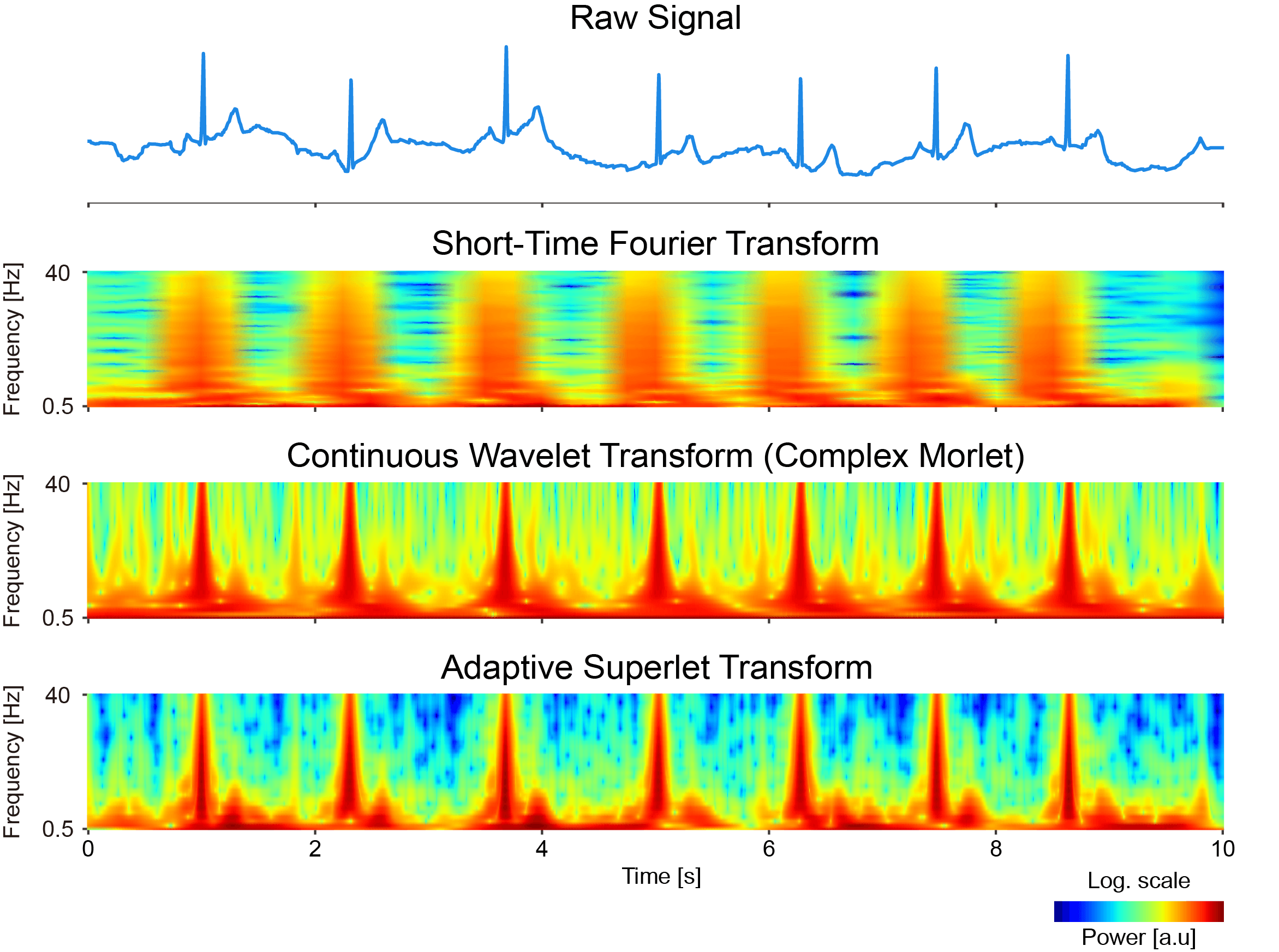}
\caption{
Comparison of time-frequency representations of a raw ECG signal using different transformation methods. From top to bottom: raw ECG, short-time Fourier transform (STFT) spectrogram, continuous wavelet transform (CWT) scalogram, and adaptive superlet transform (ASLT) scalogram. The ASLT provides sharper localization of transient components, such as QRS complexes, and effectively reduces spectral smearing and cross-band interference compared with STFT and CWT.
}
\label{fig:figure_a1}
\end{figure}

Physiological signals such as ECG and electroencephalograms (EEG) exhibit narrower frequency ranges and require high-resolution time–frequency analysis. Conventional time–frequency transformation methods, such as short-time Fourier transform (STFT) and continuous wavelet transform (CWT), are susceptible to cross-band contamination in the presence of strong spectral neighbors~\cite{ref46}. This phenomenon can degrade spectral estimation, leading to the over-representation of features such as the QRS complex in ECG signals or distortion of noise characteristics. The superlet transform~\cite{ref30} has recently emerged as a promising alternative to address these limitations. The superlet approach combines multiple wavelets with varying cycle counts at each center frequency, thereby achieving an improved trade-off between time and frequency resolutions. This method is particularly advantageous for physiological signals that require sharp and stable time–frequency representations within a limited frequency range, as visually demonstrated in the comparative results shown in Figure~\ref{fig:figure_a1}.

% figure-a1

\subsubsection*{Adaptive Superlet Transform}
We denote an ECG record as \( s \in \mathbb{R}^{T} \), where \( T \) represents the temporal length of the signal. This signal is transformed by employing the adaptive superlet transform (ASLT) to generate a scalogram. The ASLT is computed as follows: First, for a given frequency \( f \) and order \( k \), the complex Morlet wavelet is defined as
\begin{equation}
\psi_{f,k}(t) = \frac{1}{\sigma_k \sqrt{2\pi}} \, e^{i 2\pi f t} \, e^{-\frac{t^2}{2 \sigma_k^2}},
\end{equation}
where \( \sigma_k = \frac{n_k}{k_{sd} \cdot f} \), and \( n_k \) is the number of cycles for the \( k \)-th wavelet, which is determined adaptively. In multiplicative mode, the cycles are scaled as \( n_k = k \cdot n_0 \), where \( n_0 \) is the base cycle count (set to 1 in our implementation). Following the conventions of a previous study~\cite{ref30}, we fixed \( k_{sd} = 5 \) as a constant to control the bandwidth of the Gaussian envelope. The wavelet transform for a specific frequency and order is thus
\begin{equation}
W_{f,k}(t) = \sqrt{2} \cdot \int_{-\infty}^{\infty} s(\tau) \psi_{f,k}^{*}(\tau - t) \, d\tau,
\end{equation}
where \( * \) denotes the complex conjugate. The factor \( \sqrt{2} \) is included to compensate for the analytic nature of the complex Morlet wavelet, which recovers only half the power of a real-valued signal.

The ASLT aggregates these wavelet transforms across a set of orders \( k = 1, 2, \ldots, K_f \), where \( K_f \) is the frequency-dependent upper limit of the orders, computed as
\begin{equation}
K_f = o_{\text{min}} + \operatorname{round}\left( (o_{\text{max}} - o_{\text{min}}) \cdot \frac{f - f_{\text{min}}}{f_{\text{max}} - f_{\text{min}}} \right),
\end{equation}
where \( o_{\text{min}} \) and \( o_{\text{max}} \) are the minimum and maximum order limits (set to 1 and 16, respectively), and \( f_{\text{min}} \) and \( f_{\text{max}} \) the bounds of the frequency range. The final scalogram is obtained by computing the geometric mean of the wavelet transform magnitudes in the following order:
\begin{equation}
\text{ASLT}(s, f, t) = \exp\left( \frac{1}{K_f} \sum_{k=1}^{K_f} \log(W_{f,k}(t) + \epsilon) \right),
\end{equation}
where \( \epsilon = 10^{-12} \) is a small constant for numerical stability. Orders beyond \( K_f \) (up to \( o_{\text{max}} = 16 \)) are excluded to ensure that only relevant wavelet scales contribute at each frequency.

\subsubsection*{Scalogram Generation and Normalization}
For our implementation, we applied this transform over a frequency range of 0.5--40~Hz, converting the 1D ECG signal \( s \in \mathbb{R}^{T} \) into a 2D scalogram \( X \in \mathbb{R}^{F \times T} \), where \( F \) is the number of frequency bins. This frequency range was selected based on the prior physiological knowledge that the most diagnostically relevant ECG activity, including that of the PQRST complex, occurs above 0.5~Hz~\cite{ref50}. The raw scalogram dimensions were too large for direct use as model inputs. We addressed this by downsampling the scalogram along the time axis and resizing it to a manageable \( 32 \times 256 \) grid (representing 32 frequency bins and 256 time points). Subsequently, we computed the magnitude of the scalogram, squared it, and applied a logarithmic transformation to convert it into a log-scale representation, i.e., \( X_{\text{log}} = \log_{10}(|\text{ASLT}(s, f, t)|^2) \).

% figure-a2
\begin{figure}[!ht]
\centering
\includegraphics[width=\linewidth]{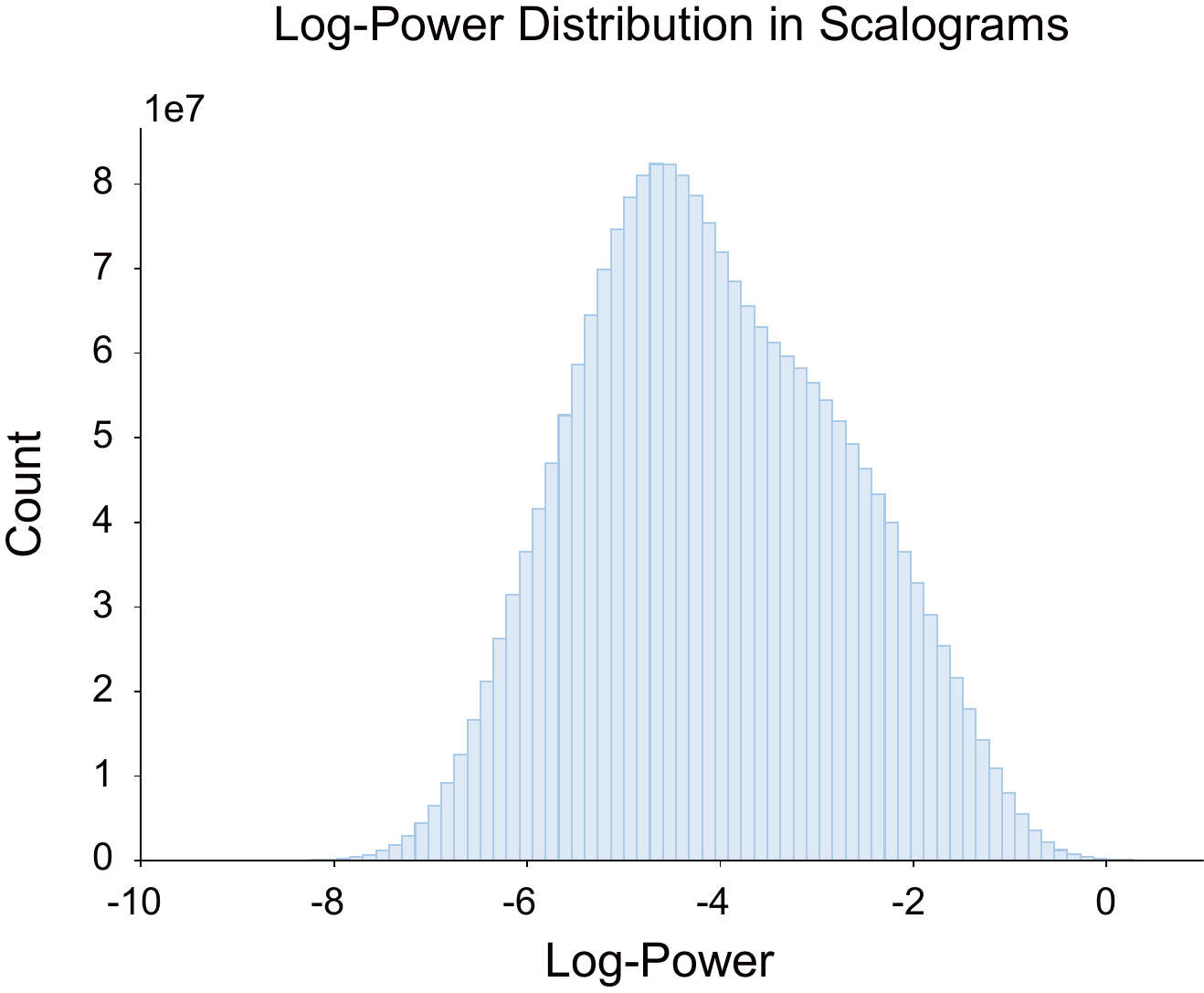}
\caption{
Histogram of log-power values in scalograms generated using the adaptive superlet transform. The distribution typically ranges between \(-8\) and \(0\), which was used for value clipping and normalization prior to model input.
}
\label{fig:figure_a2}
\end{figure}

Analysis of the resulting scalograms, as depicted in Figure~\ref{fig:figure_a2}, revealed that most values in the training set range between \(-8\) and \(0\). Accordingly, we clipped the log-scaled values to this range, i.e., \( X_{\text{clipped}} = \text{clip}(X_{\text{log}}, -8, 0) \), and linearly mapped them to the 8-bit range \([0, 255]\) to mimic grayscale image intensities
\begin{equation}
X_{\text{mapped}} = \operatorname{round} \left( 255 \cdot \frac{X_{\text{clipped}} - (-8)}{0 - (-8)} \right).
\end{equation}
This mapping compresses the dynamic range of the input while preserving essential features to stabilize training and simplify normalization. Subsequently, the mapped integer values were rescaled to the \([-1, 1]\) range via min–max normalization, i.e., \( x = 2 \cdot (X_{\text{mapped}} / 255) - 1 \), ensuring compatibility with the input requirements of the diffusion model.

\subsection*{Model Architecture}
\subsubsection*{Denoising U-Net}
The pixel-based diffusion model (DM)~\cite{ref47} and the latent diffusion model (LDM)~\cite{ref48} both employ a denoising U-Net to estimate the noise added at each diffusion time step. In the DM, the U-Net operates on high-resolution scalograms in the pixel domain, requiring additional downsampling and upsampling blocks. By contrast, LDM applies the U-Net to lower-dimensional latent representations generated by a preceding autoencoder, thereby reducing architectural complexity and computational overhead. Despite these differences, both models share the same multi-scale U-Net backbone.

\subsubsection*{Autoencoder for LDM}
The LDM framework is adapted for restoration rather than generation by employing a deterministic autoencoder instead of a variational autoencoder to preserve the fidelity of the ECG scalograms. The autoencoder is pre-trained on clean ECG data using a reconstruction loss and kept frozen during the diffusion model training to ensure stable latent representations. The encoder compresses high-level ECG features into a latent space, which the U-Net then denoises over the course of the diffusion time steps. Subsequently, the decoder reconstructs the scalogram at the original resolution. Denoising in this compact domain reduces the computational cost while preserving critical signal structure.

\subsection*{Training and Inference Strategy}
\subsubsection*{Diffusion Model Training Process}
The DM and LDM are both trained using the denoising diffusion probabilistic models (DDPM) framework~\cite{ref31}. In this approach, a forward noising process gradually adds Gaussian noise to the scalogram over \( T \) time steps, thus degrading the original structure as follows:
\begin{equation}
q(x_t | x_{t-1}) = \mathcal{N}(x_t; \sqrt{1 - \beta_t} x_{t-1}, \beta_t I),
\end{equation}
where \(\beta_t\) is a predefined noise scheduling parameter that increases with \( t \), and \( I \) is the identity matrix. The cumulative effect of this process is captured by \(\alpha_t = 1 - \beta_t\) and \(\bar{\alpha}_t = \prod_{s=1}^t \alpha_s\), where \(\sqrt{\bar{\alpha}_t}\) quantifies the proportion of the original scalogram \( x_0 \) retained at timestep \( t \).

The entire forward process can be expressed as
\begin{equation}
q(x_t | x_0) = \mathcal{N}(x_t; \sqrt{\bar{\alpha}_t} x_0, (1 - \bar{\alpha}_t) I).
\end{equation}
This process can also be expressed equivalently as
\begin{equation}
x_t = \sqrt{\bar{\alpha}_t} x_0 + \sqrt{1 - \bar{\alpha}_t} \epsilon, \quad \epsilon \sim \mathcal{N}(0, I),
\end{equation}
where \( x_0 \) is the initial scalogram, and \(\epsilon\) is random Gaussian noise.

The model learns the reverse diffusion process to denoise \( x_t \) step-by-step as follows:
\begin{equation}
p_{\theta}(x_{t-1} | x_t) = \mathcal{N}(x_{t-1}; \mu_{\theta}(x_t, t), \tilde{\beta}_t I),
\end{equation}
where \(\mu_{\theta}(x_t, t)\) is the mean estimated by the denoising U-Net parameterized by \(\theta\), and the variance is defined as
\begin{equation}
\tilde{\beta}_t = \frac{1 - \bar{\alpha}_{t-1}}{1 - \bar{\alpha}_t} \beta_t.
\end{equation}

During training, we optimize a simplified objective to predict the added noise \(\epsilon\) at each time step \( t \)
\begin{equation}
\mathcal{L} = \mathbb{E}_{x_0, \epsilon, t} \left[ || \epsilon - \epsilon_{\theta}(x_t, t) ||^2 \right],
\end{equation}
where \(\epsilon_{\theta}(x_t, t)\) is the noise predicted by the model. In the case of the LDM, the input scalograms are first encoded into latent representations using a pre-trained autoencoder, after which the forward and reverse diffusion processes are executed in this compressed domain. This approach reduces the computational overhead while preserving scalogram fidelity.

\subsubsection*{Inference with DDPM and DDIM}
After training the diffusion model on clean ECG data, we reconstruct the scalograms using different sampling strategies. Rather than running the entire diffusion process over \( T \) steps, we limit it to a smaller number \( \lambda \leq T \), enabling analysis of ECG noise sensitivity as a function of \(\lambda\). The reconstruction error between the input scalogram and the denoised output serves as an anomaly score, with larger errors indicating higher noise levels.

\paragraph{DDPM.} The original DDPM sampling method follows a fully Markovian reverse process, iterating over all \( T \) timesteps (\( t = T, T-1, \dots, 1 \)) to gradually refine the image. The sampling step is
\begin{equation}
x_{t-1} = \frac{1}{\sqrt{\alpha_t}} \left( x_t - \frac{1 - \alpha_t}{\sqrt{1 - \bar{\alpha}_t}} \epsilon_{\theta}(x_t, t) \right) + \sqrt{\tilde{\beta}_t} \epsilon_t, \quad \epsilon_t \sim \mathcal{N}(0, I),
\end{equation}
where \( \epsilon_t \) is an independent Gaussian noise term added at each step. Although this method produces high-quality reconstructions, it is computationally expensive owing to the numerous iterations.

\paragraph{DDIM.} To improve efficiency, we employ the denoising diffusion implicit models (DDIM)~\cite{ref32} approach, which allows fewer sampling steps (e.g., with a stride of 10) by introducing a non-Markovian, deterministic sampling process. The DDIM sampling step is as follows:
\begin{equation}
\begin{split}
x_{t-1} &= \sqrt{\bar{\alpha}_{t-1}} \left( \frac{x_t - \sqrt{1 - \bar{\alpha}_t} \cdot \epsilon_{\theta}(x_t, t)}{\sqrt{\bar{\alpha}_t}} \right) \\
&\quad + \sqrt{1 - \bar{\alpha}_{t-1} - \sigma_t^2} \cdot \epsilon_{\theta}(x_t, t) + \sigma_t \epsilon_t, \quad \epsilon_t \sim \mathcal{N}(0, I).
\end{split}
\end{equation}

In our implementation, we set \(\sigma_t = 0\), which eliminates the stochastic term \(\sigma_t \epsilon_t\) and makes the process fully deterministic. This deterministic formulation enhances computational efficiency and facilitates rapid denoising, making it suitable for real-time ECG noise quantification in clinical or wearable settings.

\subsection*{Model Evaluation and Refinement Strategy}
We hypothesized that a model trained exclusively on clean data would produce significant discrepancies between the input and output when applied to noisy ECG signals. As such, we employed pointwise and distribution-level metrics to systematically assess how effectively each model quantifies ECG noise.

\subsubsection*{Image-Based Evaluation Metrics}
We first considered the following three standard image-based evaluation metrics that quantify the differences between the input and reconstruction:
\begin{itemize}
\item \textbf{Mean Absolute Error (MAE)}: This measures the average absolute pixel difference between the input and reconstruction, quantifying the overall reconstruction accuracy.
\item \textbf{Structural Similarity Index Measure (SSIM)}~\cite{ref49}: This metric evaluates perceptual quality by comparing the luminance, contrast, and structure between original and denoised scalograms. A low SSIM score indicates significant structural distortions due to noise or artifacts.
\item \textbf{Peak Signal-to-Noise Ratio (PSNR)}: This computes the ratio between the maximum possible signal power and reconstruction error power. Higher PSNR values reflect better image fidelity.
\end{itemize}

Although these image-based metrics provide useful pointwise measures of reconstruction quality, they do not directly assess whether the model effectively separates clean from noisy inputs at the distribution level. Moreover, because human-provided noise annotations can be subjective and inconsistently applied, particularly near the boundary between clean and noisy signals, direct evaluation against such labels may lead to misleading conclusions.

\subsubsection*{Distributional Evaluation via Wasserstein-1 Distance}
We adopt a distributional perspective to address these limitations, assuming that an effective model should clearly separate the reconstruction error distributions of clean and noisy signals. Based on this assumption, we quantify model performance using the Wasserstein-1 distance ($W_1$) as follows:
\begin{equation}
W_1(P, Q) = \inf_{\gamma \in \Pi(P, Q)} \mathbb{E}_{(x, y) \sim \gamma} [ ||x - y|| ],
\end{equation}
where \( P \) and \( Q \) denote the reconstruction error distributions for clean and noisy ECG samples, respectively, and \( \Pi(P, Q) \) represents the set of all joint distributions with marginals \( P \) and \( Q \). A higher \( W_1 \) value reflects a stronger separation between clean and noisy reconstruction error distributions, indicating more effective noise discrimination.

\subsubsection*{Comparative Evaluation of Configurations}
Following the introduction of \( W_1 \) as the primary evaluation metric, we conducted a systematic comparison across various configurations, including model architecture (DM vs.\ LDM), sampling strategy (DDPM vs.\ DDIM), diffusion timestep (\( \lambda \)), and image-based metrics (MAE, SSIM, and PSNR). For each configuration, the ability of the model to discriminate between clean and noisy ECG signals was assessed. This was quantified using  \( W_1 \) as a measure of distributional divergence in reconstruction errors.

Before computing \( W_1 \), the reconstruction error distributions were standardized to a zero mean and unit variance. This standardization ensures that the computed distance reflects differences in the distributional shape rather than scale, enabling a fair comparison across configurations.

\subsubsection*{Training Set Refinement and Model Retraining}
Based on a systematic evaluation, we selected the configuration that produced the largest distributional divergence between the clean and noisy ECGs. Using the model trained with this configuration, we refined the clean-labeled dataset by identifying and removing potentially mislabeled samples. The model was then retrained on this curated subset to further improve its robustness. This refinement enhances the ability of the model to learn in-distribution representations by restricting training to consistently clean signals, thereby improving its sensitivity to out-of-distribution noise.

%%%%%%%%%%%%%%%%%%%%%%%%%%%%%%%%%%%%%%%%%%%%%%%%%%%%%%%%%%%%%%%%%%%%%%%%%%%%%%

\section*{Results}

\subsection*{PTB-XL Dataset}
Our approach to diffusion-based ECG noise quantification was evaluated using the PTB-XL dataset~\cite{ref14}, which includes 21,799 twelve-lead ECG recordings from 18,869 patients. Each 10-s recording was provided at both 100 and 500~Hz; we selected the 500~Hz version for training and evaluation owing to its higher temporal resolution. Noise annotations are provided at the lead level and categorize noise into baseline drift, static noise, burst noise, and electrode problems. Multiple types of noise may be present in a single recording. Owing to the limited sample size and potential for metric distortion, the electrode problem category was excluded from the analysis. A summary of the number of clean and noise-labeled leads in the dataset is presented in Table~\ref{tab:table_a1}.

% table-a1
\begin{table}[!ht]
\begin{tabular}{@{}cccccc@{}}
\toprule
Split & \# Clean & \multicolumn{4}{c}{\# Noise} \\ \cmidrule(l){3-6} 
 &
   &
  \begin{tabular}[c]{@{}c@{}}Static \\ Noise\end{tabular} &
  \begin{tabular}[c]{@{}c@{}}Burst \\ Noise\end{tabular} &
  \begin{tabular}[c]{@{}c@{}}Baseline \\ Drift\end{tabular} &
  \begin{tabular}[c]{@{}c@{}}Electrode \\ Problems\end{tabular} \\ \midrule
Train & 209781   & -      & -     & -     & -   \\
Test  & 23369    & 21795  & 3098  & 4569  & 57  \\ \bottomrule
\end{tabular}
\caption{
Summary of clean and noisy ECG sample counts in the PTB-XL dataset. All counts are reported per lead.
}
\label{tab:table_a1}
\end{table}

% figure-2
\begin{figure*}[!ht]
\centering
\includegraphics[width=\linewidth]{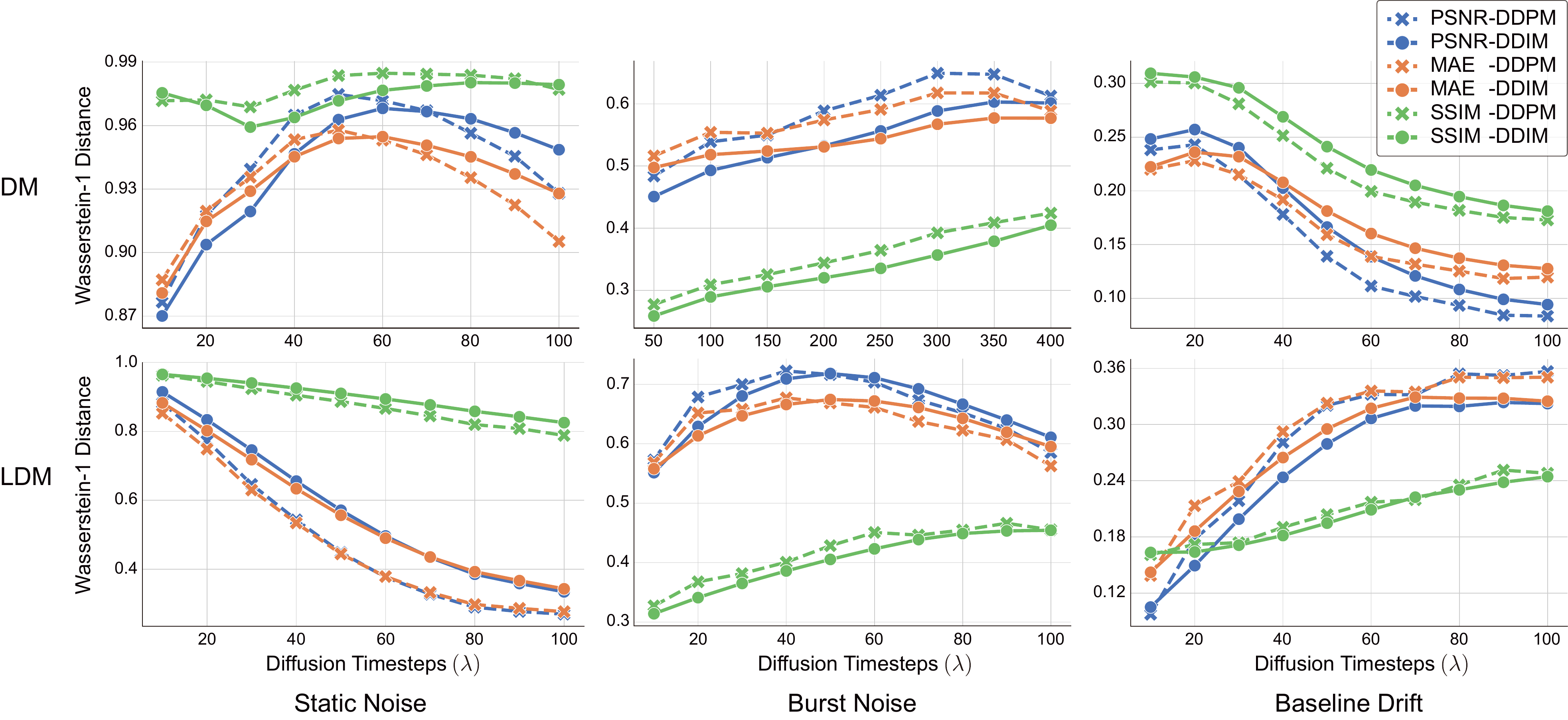}
\caption{
Quantitative comparison of ECG noise discrimination performance using Wasserstein-1 distance ($W_1$). Each subplot shows the $W_1$ between clean and noisy ECG distributions for static noise, burst noise, and baseline drift, evaluated across various model configurations: diffusion model type (DM = pixel-based diffusion model, LDM = latent diffusion model), sampling strategy (DDPM = denoising diffusion probabilistic model, DDIM = denoising diffusion implicit model), image-based metric (PSNR = peak signal-to-noise ratio, MAE = mean absolute error, SSIM = structural similarity index), and diffusion time step ($\lambda$).
}
\label{fig:figure2}
\end{figure*}

% table-1
\begin{table*}[!ht]
\centering
\begin{tabular}{@{}llcccccc@{}}
\toprule
Model & Noise type & \multicolumn{6}{c}{Wasserstein-1 distance}                                     \\ \cmidrule(l){3-8} 
      &            & \multicolumn{3}{c}{DDPM}               & \multicolumn{3}{c}{DDIM}           \\ \cmidrule(l){3-5}\cmidrule(l){6-8} 
      &            & PSNR        & MAE         & SSIM       & PSNR       & MAE        & SSIM        \\ \midrule
    & Static & \underline{0.975 (50)}        & 0.958 (50)  & {\textbf{0.985 (60)}} & 0.968 (60)      & 0.955 (60)  & 0.980 (80)                \\
DM  & Burst  & {\textbf{0.649 (300)}} & 0.618 (300) & 0.424 (400)               & 0.603 (350)         & 0.577 (350) & 0.405 (400)               \\
    & Baseline   & 0.243 (20)  & 0.228 (20)  & 0.301 (10) & 0.257 (20) & 0.236 (20) & 0.309 (10)  \\ \midrule
    & Static & 0.882 (10)                 & 0.853 (10)  & 0.963 (10)                & \underline{0.914 (10)} & 0.883 (10)  & {\textbf{0.965 (10)}} \\ 
LDM & Burst  & {\textbf{0.722 (40)}}  & 0.677 (40)  & 0.466 (90)                & 0.718 (50)          & 0.674 (50)  & 0.454 (100)               \\
    & Baseline   & 0.357 (100) & 0.351 (100) & 0.251 (90) & 0.324 (90) & 0.329 (70) & 0.244 (100) \\ \bottomrule
\end{tabular}
\caption{
Best-performing configurations for each ECG noise type. The table summarizes the highest Wasserstein-1 distance ($W_1$) scores achieved for static noise, burst noise, and baseline drift across combinations of model architecture (DM = pixel-based diffusion model, LDM = latent diffusion model), sampling strategy (DDPM = denoising diffusion probabilistic model, DDIM = denoising diffusion implicit model), image-based metric (PSNR = peak signal-to-noise ratio, SSIM = structural similarity index, MAE = mean absolute error), and diffusion time step $\lambda$ (shown in parentheses).
}
\label{tab:table1}
\end{table*}

% figure-3
\begin{figure*}[!ht]
\centering
\includegraphics[width=\linewidth]{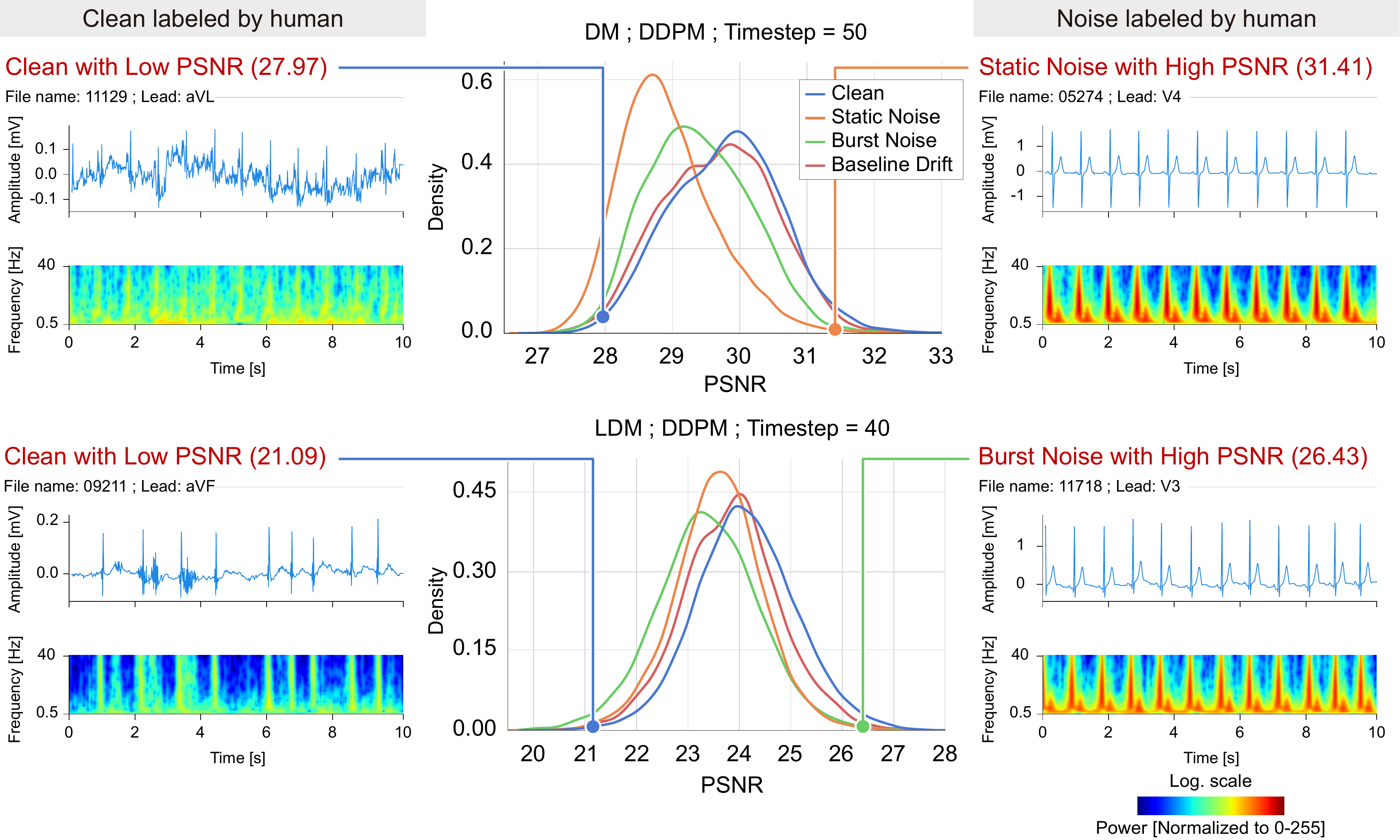}
\caption{
Representative ECG segments illustrating mismatches between model assessments based on peak signal-to-noise ratio (PSNR) and human-provided noise labels. High-PSNR segments labeled as noisy and low-PSNR segments labeled as clean suggest potential annotation inconsistencies.
}
\label{fig:figure3}
\end{figure*}

\subsection*{Comparative Noise Discrimination Analysis}
We evaluated the discriminative performance of our models across three annotated noise types by measuring the $W_1$ between clean and noisy signal distributions. The results are summarized in Figure~\ref{fig:figure2}, which provides a comprehensive comparison across different model architectures, sampling strategies, diffusion steps ($\lambda$), and reconstruction error metrics.

The best-performing configurations are listed in Table~\ref{tab:table1}. For static noise, the optimal setup was (DM, DDPM, SSIM, $\lambda=60$), yielding $W_1 = 0.985$, followed closely by (DM, DDPM, PSNR, $\lambda=50$) with $W_1 = 0.975$. For burst noise, the top-performing configuration was (LDM, DDPM, PSNR, $\lambda=40$), achieving $W_1 = 0.722$. By contrast, baseline drift exhibited minimal separability from clean signals, with the highest score observed for (LDM, DDPM, PSNR, $\lambda=100$) at only $W_1 = 0.357$. This limited separability is likely because baseline drift primarily occurs below 0.5~Hz, a frequency range that was deliberately excluded in our ASLT configuration.

Given its consistently high $W_1$ scores for static and burst noises, PSNR was selected as the primary metric for subsequent noise quantification analyses. The LDM demonstrated greater efficiency than the DM, achieving optimal $W_1$ scores and requiring fewer diffusion steps. Static noise, characterized by widespread distortion, was effectively identified at shallower diffusion levels. By contrast burst noise, which is localized and transient, required deeper denoising to be adequately captured by the PSNR, which averages the differences across the entire scalogram.

\subsection*{Analysis of Label Consistency}
We conducted a PSNR-based analysis using the best configurations for static noise (DM, DDPM, $\lambda=50$) and burst noise (LDM, DDPM, $\lambda=40$) to assess potential inconsistencies in human-provided annotations. As shown in Figure~\ref{fig:figure3}, the noise-labeled samples with a high PSNR appeared visually clean, whereas the clean-labeled samples with a low PSNR exhibited clear noise artifacts. These observations suggest potential mismatches between labels and the actual signal quality, which motivated the refinement of the training dataset.

% figure-4
\begin{figure*}[!ht]
\centering
\includegraphics[width=\linewidth]{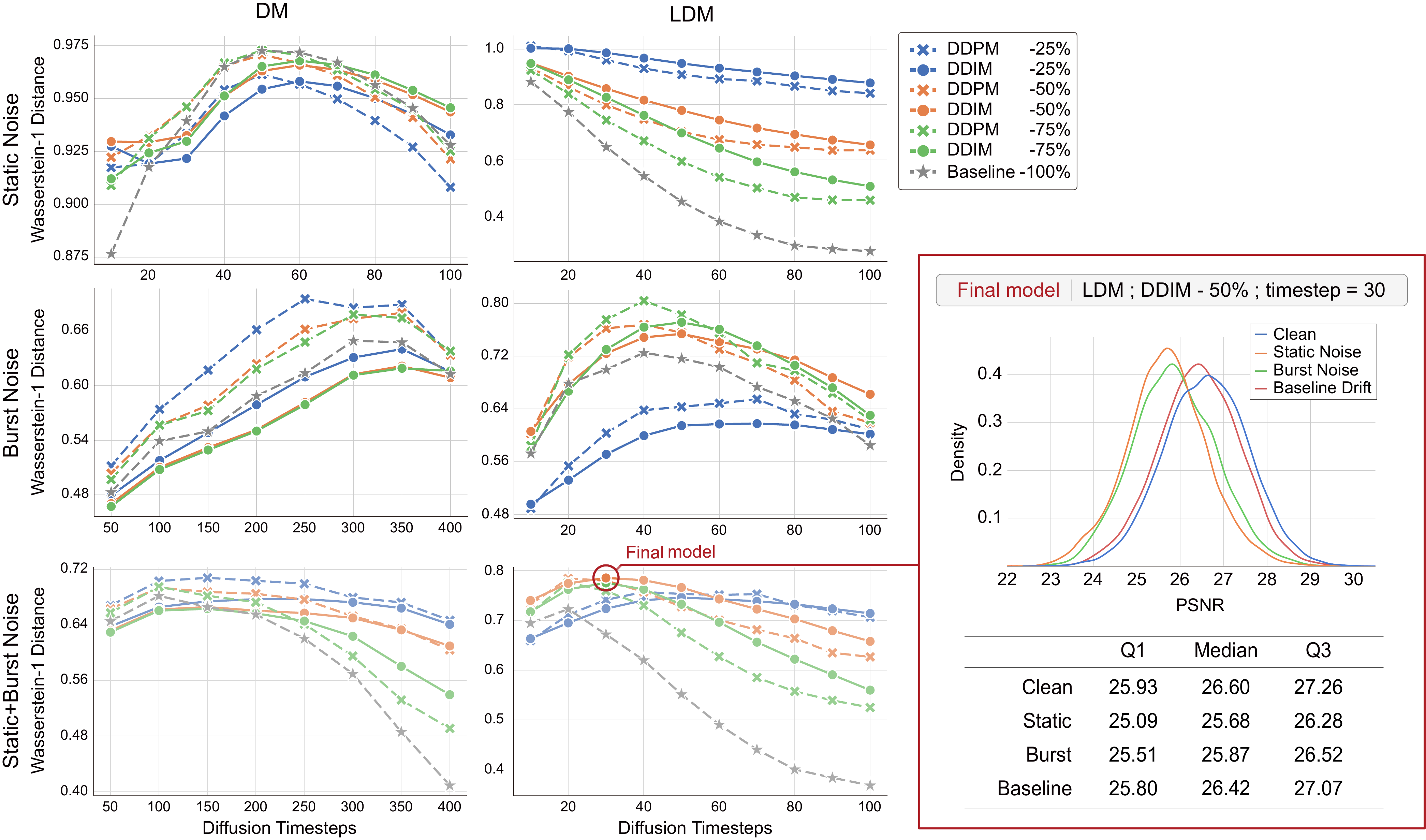}
\caption{
Noise quantification performance of the pixel-based diffusion model (DM) and latent diffusion model (LDM) after training set refinement. Results are shown for varying refinement thresholds (\(N = 25\%, 50\%, 75\%\)). The final model, LDM with denoising diffusion implicit model (DDIM) sampling and \(\lambda=30\) diffusion steps, was selected based on the highest arithmetic mean of Wasserstein-1 distance ($W_1$) scores for static and burst noise.
}
\label{fig:figure4}
\end{figure*}

\subsection*{Dataset Refining and Model Retraining}
Building on this analysis, we refined the training set to improve model robustness by filtering out potentially mislabeled samples. Clean ECG signals were selected for retraining based on the intersection of the top \(N\)\% PSNR-ranked samples from (DM, DDPM, \(\lambda=50\)) and (LDM, DDPM, \(\lambda=40\)) outputs, formally defined as \(\mathcal{D}_{\text{DM-N\%}} \cap \mathcal{D}_{\text{LDM-N\%}}\). We conducted experiments with \(N = 25\%\), \(50\%\), and \(75\%\) subsets and compared them with the baseline models trained on the original training set. As shown in Figure~\ref{fig:figure4}, the DM exhibited negligible improvements from this refinement, whereas the LDM showed substantial gains in static noise quantification. Considering both performance and computational efficiency, (LDM, DDIM, \(\lambda=30\)) was ultimately selected as the final model configuration.

\subsection*{Validation on External Datasets}
We assessed the generalizability of our final noise quantification model by validating its performance on three external ECG datasets: the Brno University of Technology ECG Quality Database (BUT QDB)~\cite{ref23}, PhysioNet/Computing in Cardiology Challenge 2011 dataset (CinC Challenge 2011)~\cite{ref24}, and MIT-BIH Noise Stress Test Database (NSTDB)~\cite{ref25}. To ensure consistency with the PTB-XL protocol, all external recordings were segmented into 10-second intervals, and each segment was independently evaluated using the model. The results are summarized below.

\subsubsection*{BUT QDB Dataset}
The BUT QDB comprises single-lead ECG recordings collected during daily activities, sampled at 1,000~Hz, and annotated into three quality classes: Class~1 (clearly visible signals), Class~2 (moderate noise with a recognizable QRS complex), and Class~3 (severely contaminated and unsuitable for analysis). The distribution of PSNR values across these classes is presented as a histogram in Figure~\ref{fig:figure5}(a). The results exhibit a clear decreasing trend in the PSNR values: Class~1 $>$ Class~2 $>$ Class~3. Notably, the distributions for Classes~1 and ~3 show minimal overlap, indicating that the model effectively distinguishes between clean and heavily corrupted signals.

\subsubsection*{CinC Challenge 2011 Dataset}
The CinC Challenge 2011 dataset comprises 998 twelve-lead ECG recordings labeled as either ``acceptable'' or ``unacceptable,'' with each lead sampled at 500~Hz. However, the dataset does not specify which individual leads are affected by noise, limiting its utility for lead-level noise detection. Falk et al.~\cite{ref10} addressed this issue by synthetically assigning clean or noisy labels to individual leads. Using these re-annotated labels, we analyzed the PSNR distributions for clean versus noisy segments.

Our model exhibited a substantial distributional overlap in contrast to Falk et al., who reported near-separable distributions between the two classes using their modulation spectral-based quality index (MS-QI). We investigated this discrepancy by extracting samples from the overlapping region and visually comparing the clean- and noise-labeled recordings, as shown in Figure~\ref{fig:figure5}(b). The comparison revealed that these samples exhibit similar noise levels.

\subsubsection*{NSTDB Dataset}
\begin{figure*}[!ht]
\centering
\includegraphics[width=\linewidth]{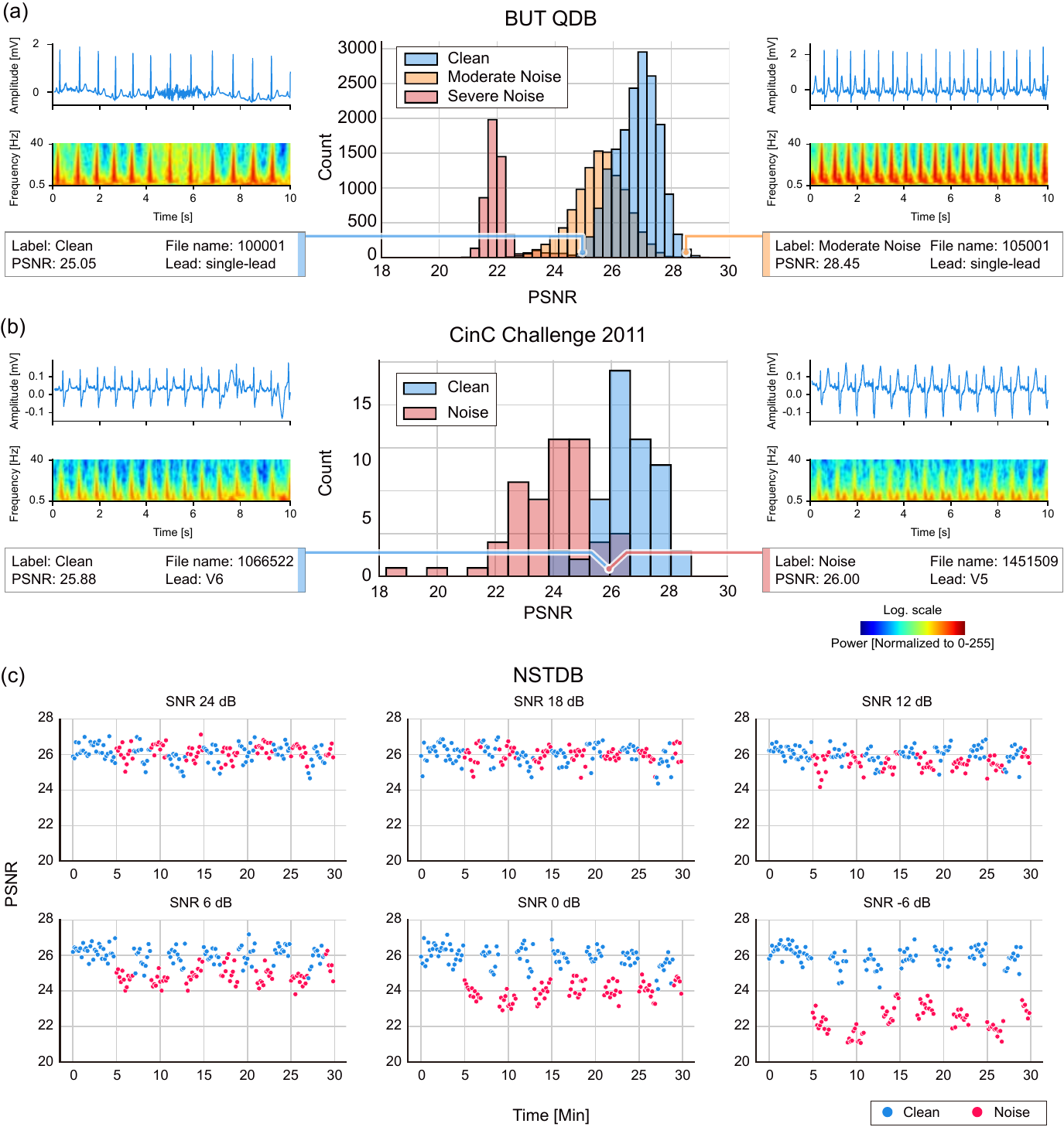}
\caption{
Evaluation of the proposed noise quantification model on three external ECG datasets. 
(a) Peak signal-to-noise ratio (PSNR) distributions across quality classes in the Brno University of Technology ECG Quality Database (BUT QDB), demonstrating clear separation between clean and severely noisy signals. 
(b) PSNR distributions for re-annotated leads in the PhysioNet/Computing in Cardiology (CinC) Challenge 2011 dataset, highlighting distributional overlap and the limitations of binary labeling. 
(c) PSNR profiles from the MIT-BIH Noise Stress Test Database (NSTDB) across varying signal-to-noise ratio (SNR) levels, illustrating the sensitivity of the model to graded noise severity.
}
\label{fig:figure5}
\end{figure*}

The NSTDB comprises ECG recordings from two subjects, with artificial noise injected into previously clean signals. The data comprising two-lead ECGs were digitized at 360~Hz. Each recording includes 5 min of clean signal, followed by alternating 2-min segments of noise-injected and clean data. Noise levels vary across six signal-to-noise ratio (SNR) conditions, ranging from 24~dB to --6~dB. Figure~\ref{fig:figure5}(c) shows the PSNR outputs, with the noise-injected and clean segments shown in red and blue, respectively. The clean and noisy segments are clearly separable at an SNR of --6~dB. However, as the SNR increases (e.g., approaching 24~dB), the distinction becomes less pronounced, reflecting lower noise intensity.

\subsection*{Quantitative Comparison with Existing Methods}
We conducted a comparative analysis against a range of established SQIs and a recently proposed deep learning-based method to further evaluate the performance of the proposed diffusion-based noise quantification model. Table~\ref{tab:table2} lists the $W_1$ scores achieved by each method across three benchmark datasets: PTB-XL, BUT QDB, and CinC Challenge 2011. Across all datasets and noise categories, the diffusion-based model outperformed conventional SQIs such as qSQI, kSQI, and pSQI, as well as the CNN-LSTM learning-based method. The model achieved the highest $W_1$ scores in three of the five evaluation settings and ranked second in the remaining two. Most notably, our model recorded a macro-average $W_1$ score of 1.308, substantially exceeding that of CNN-LSTM (0.884), the next-best performer, by over 48\%. 

Although $W_1$ serves as a robust distribution metric, it may be sensitive to labeling inconsistencies in small datasets. In particular, for the CinC Challenge 2011 dataset, which contains only 138 re-annotated leads, performance comparisons using $W_1$ may be less reliable owing to the limited sample size and potential label noise.

% table-2
\begin{table*}[!ht]
\centering
\begin{tabular}{@{}ccccccc@{}}
\toprule
Method     & \multicolumn{2}{c}{PTB-XL}       & \multicolumn{2}{c}{BUT QDB}     & CinC Challenge 2011 & Macro Avg. \\ 
\cmidrule(l){2-3} \cmidrule(l){4-5} \cmidrule(l){6-6}
           & Static Noise   & Burst Noise    & Moderate Noise & Severe Noise   & Noise               &             \\ \midrule
qSQI~\cite{ref28}         & \underline{0.483} & 0.089 & 0.441 & 1.314 & 0.346 & 0.535 \\
kSQI~\cite{ref64}         & 0.176 & 0.111 & \underline{0.678} & 1.565 & 0.616 & 0.629 \\
pSQI~\cite{ref65}         & 0.029 & 0.246 & 0.404 & 2.076 & 0.202 & 0.591 \\
basSQI~\cite{ref11}       & 0.092 & 0.161 & 0.424 & 1.248 & 1.052 & 0.595 \\
Fuzzy~\cite{ref28}        & 0.303 & 0.049 & 0.235 & 1.817 & 0.758 & 0.632 \\
MS-QI~\cite{ref10}        & 0.405 & \underline{0.404} & 0.182 & 1.544 & \textbf{1.758} & 0.859 \\
wPMF~\cite{ref29}         & 0.083 & 0.100 & 0.399 & 1.538 & 1.348 & 0.694 \\
AverageQRS~\cite{ref27}   & 0.023 & 0.075 & 0.231 & 2.088 & 0.361 & 0.556 \\
CNN-LSTM~\cite{ref43}     & 0.127 & 0.226 & 0.477 & \textbf{2.313} & 1.277 & \underline{0.884} \\ \midrule
\textbf{Diffusion}        & \textbf{0.857} & \textbf{0.724} & \textbf{1.271} & \underline{2.221} & \underline{1.467} & \textbf{1.308} \\ \bottomrule
\end{tabular}
\caption{
Comparison of ECG noise discrimination performance across various signal quality indices (SQIs) and learning-based methods. The evaluation is based on the Wasserstein-1 distance ($W_1$) between clean and noisy signal distributions across five conditions from PTB-XL, BUT QDB, and CinC Challenge 2011. The rightmost column shows the macro average across all five conditions. The numbers in bold indicate the highest $W_1$ score for each condition, and the underlined values indicate the second highest. The proposed diffusion-based model achieves the highest or second-highest $W_1$ in all settings.
}
\label{tab:table2}
\end{table*}

% figure-5

% figure-6
\begin{figure*}[!ht]
\centering
\includegraphics[width=\linewidth]{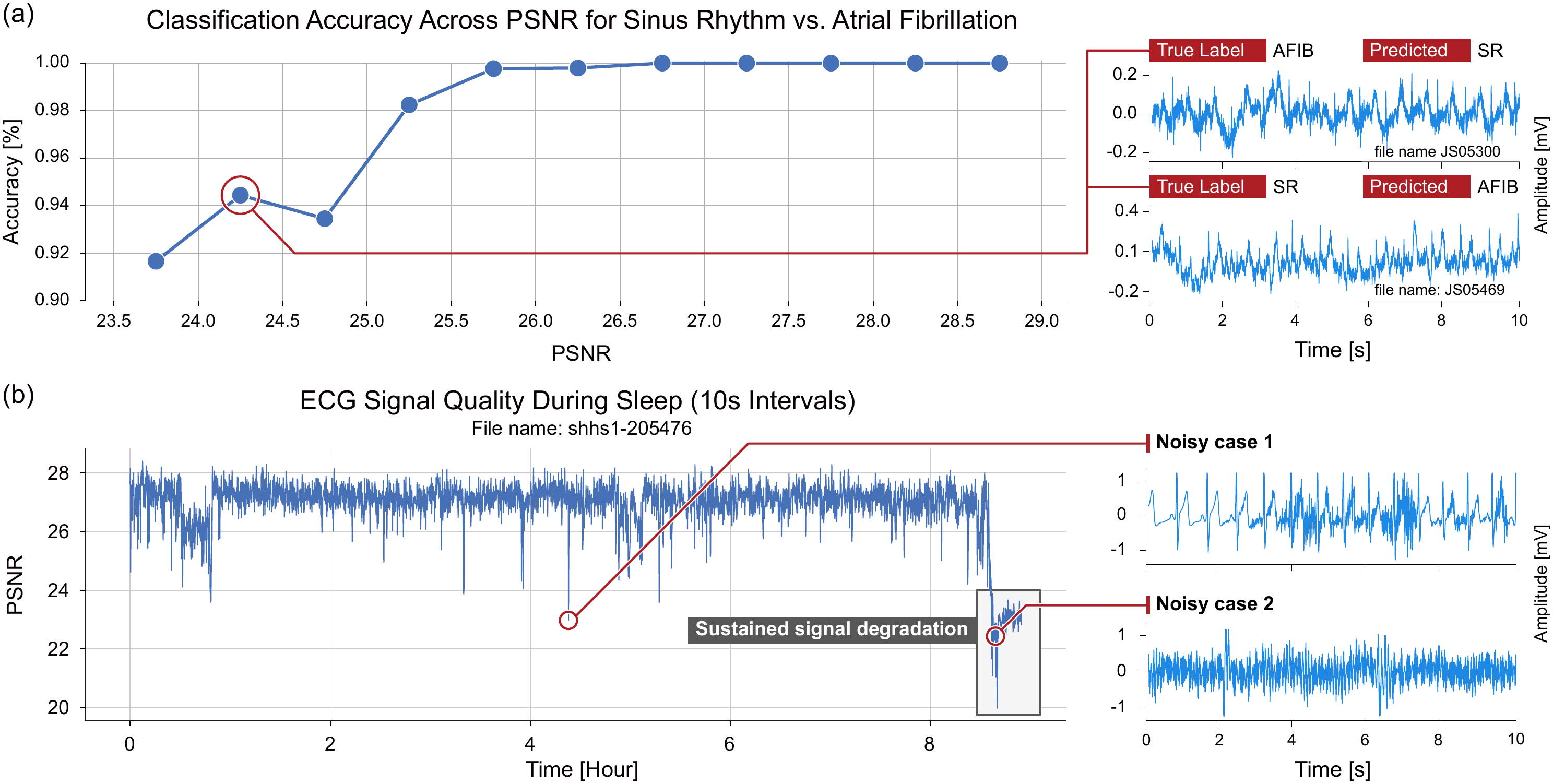}
\caption{
Evaluation of the clinical and real-world applications of the proposed noise quantification model. 
(a) Classification accuracy across peak signal-to-noise ratio (PSNR) levels for detecting sinus rhythm versus atrial fibrillation in the Arrhythmia dataset, highlighting the impact of signal quality on diagnostic reliability. 
(b) Long-term ECG monitoring using the Sleep Heart Health Study (SHHS) dataset, where sustained PSNR degradation corresponds to prolonged signal failure.
}
\label{fig:figure6}
\end{figure*}

\subsection*{Clinical and Real-World Applications}
We demonstrated the practical applicability of our model by conducting two additional experiments targeting clinical classification and long-term signal monitoring.

\subsubsection*{Effect of Signal Quality on Arrhythmia Classification}
We first assessed the clinical relevance of our model using the Arrhythmia dataset~\cite{ref34}, which comprises 45,152 twelve-lead ECG samples recorded at 500~Hz, each lasting 10 s. For this experiment, we extracted lead~II signals and performed binary classification to distinguish between sinus rhythm (SR) and atrial fibrillation (AFIB). The dataset contains 8,125 SR and 1,780 AFIB instances. The data were randomly split into training and test sets at a 4:1 ratio, and a classifier was trained on the training subset. Figure~\ref{fig:figure6}(a) shows the classification accuracy stratified by PSNR values. The results reveal a clear trend: lower PSNR scores, which indicate reduced signal quality, are associated with decreased classification performance. 

\subsubsection*{Long-Term Signal Degradation Detection in SHHS}
We further evaluated the capacity of the model for long-term ECG quality monitoring using the Sleep Heart Health Study (SHHS) dataset~\cite{ref35}. Collected through overnight polysomnography, this dataset includes various physiological signals, such as EEG, ECG, and chin EMG. Given the extended recording durations typical of sleep studies, manual ECG quality assessment is infeasible. We selected one subject and segmented the ECG signal into 10-s intervals to simulate a realistic monitoring scenario. The PSNR was computed for each segment to track signal quality over time. As shown in Figure~\ref{fig:figure6}(b), the PSNR dropped below 24 after approximately 8.6 h and remained consistently low thereafter. Visual inspection confirmed that this decline coincided with a period during which cardiac activity was no longer properly captured.

\subsection*{Lead-Specific and Clinical Stratification Analysis}
We conducted analyses using the PTB-XL and CinC Challenge 2011 datasets to investigate how the signal quality varies across ECG leads and clinical contexts.

\subsubsection*{Lead-Wise Signal Quality Trends}
The results of this analysis are presented in Figure~\ref{fig:figure7}. In the PTB-XL training set, limb leads (I–III) and augmented limb leads (aVR–aVF) were excluded more frequently than precordial leads (V1–V6) when only the top 50\% of the clean-labeled data ranked by PSNR were retained. A similar trend was observed in the clean-labeled portion of the CinC Challenge 2011 dataset, where precordial leads appeared more frequently than limb or augmented limb leads. Lead-wise PSNR values further confirmed this pattern, consistently showing the following hierarchy: precordial leads > limb leads > augmented limb leads.

\subsubsection*{Clinical Condition-Based PSNR Distributions}
The diagnostic annotations in the PTB-XL dataset enabled a comparison of the PSNR values across three clinical categories: clean signals with normal cardiac function (clean-norm), clean signals with abnormalities (clean-abnormal), and noisy signals (noise). As shown in Figure~\ref{fig:figure7}(b), clean-norm signals consistently exhibited higher PSNR values than clean-abnormal signals, whereas noisy signals showed the lowest PSNR overall.

\subsection*{Limitations of Segment-Level PSNR}
Although PSNR proved effective for overall noise quantification, it has limitations when applied to entire 10-s ECG segments. Because it averages over the entire duration, short-duration, localized noise may be underestimated. As shown in Figure~\ref{fig:figure8}, a segment with localized noise had a global PSNR of 26.43, whereas the PSNR over the affected region dropped to 22.08, revealing a clear local degradation.

%%%%%%%%%%%%%%%%%%%%%%%%%%%%%%%%%%%%%%%%%%%%%%%%%%%%%%%%%%%%%%%%%%%%%%%%%%%%%%

\section*{Discussion}
This study proposes a novel framework for ECG noise quantification grounded in diffusion-based reconstruction. Unlike conventional approaches that consider noise as a classification or denoising problem and are thus inherently reliant on predefined labels or synthetic data augmentations, the proposed framework models noise as a deviation from learned representations of clean ECG morphology. This approach enables signal quality estimation through reconstruction errors without requiring explicit noise labels during inference. Across multiple public datasets and noise conditions, the proposed method demonstrated robust generalization and adequate differentiation between clean and noisy segments, thereby supporting its potential as a practical alternative to conventional noise assessment techniques. This conceptual shift, i.e., from noise labeling to recognizing anomalous deviations, warrants a detailed discussion. 

\subsection*{Technical Significance}
Conventional approaches to ECG noise assessment typically rely on predefined noise categories or explicit modeling of specific artifact types (e.g., baseline drift, motion-related interference, and electrode detachment). Although earlier methods have proven effective in controlled settings, they often struggle when faced with unmodeled, mixed, or ambiguous artifacts that fall outside their assumptions. The proposed diffusion-based reconstruction framework addresses this challenge by reframing noise quantification as a statistical deviation problem rather than a classification issue.

Unlike conventional methods such as Kalman filtering~\cite{ref2}, wavelet transforms~\cite{ref3}, empirical mode decomposition~\cite{ref4}, and deep learning-based approaches utilizing synthetic noise~\cite{ref5,ref6,ref7}, our model leverages the flexible reconstruction capabilities inherent in diffusion models. By focusing on statistical regularity rather than specific noise types, the diffusion model can generalize to a broader range of signal disruptions. Using $W_1$ as a distributional evaluation metric further enhances this flexibility, enabling model selection and refinement without direct dependence on subjective labels. This ability to evaluate model performance with reduced reliance on subjective labels is particularly important, given the inconsistencies observed in human annotations (Figure~\ref{fig:figure3}), which limit the reliability of binary or categorical labels. Continuous quality metrics such as PSNR address this issue by offering more granular and objective assessments of signal fidelity, thereby supporting a more precise identification of mislabeled or ambiguous ECG segments. Together, these objective metrics enhance the ability of our framework to effectively discriminate noise, as demonstrated by our comparative analysis (Table~\ref{tab:table2}), which showed the highest or second-highest $W_1$ scores across diverse noise conditions and datasets. This underscores the robustness and generalizability of our approach.

Generative modeling is often viewed as computationally expensive and undesirable for analyzing continuous data streams. This conceptual disadvantage may explain the slow adaptation of generative learning techniques to physiological signals analyses~\cite{ref57}. However, after rigorous investigations to determine the best combination (Table~\ref{tab:table1} and Figure~\ref{fig:figure2}), the final model based on LDM and DDIM yielded optimal performances through only three reverse diffusion steps (Figure~\ref{fig:figure4}), demonstrating the applicability of generative modeling to physiological signal analysis tasks. The notable reduction in computational steps makes it particularly suitable for real-time ECG monitoring scenarios where computational efficiency and rapid inference are crucial.

% figure-7
\begin{figure*}[!ht]
\centering
\includegraphics[width=\linewidth]{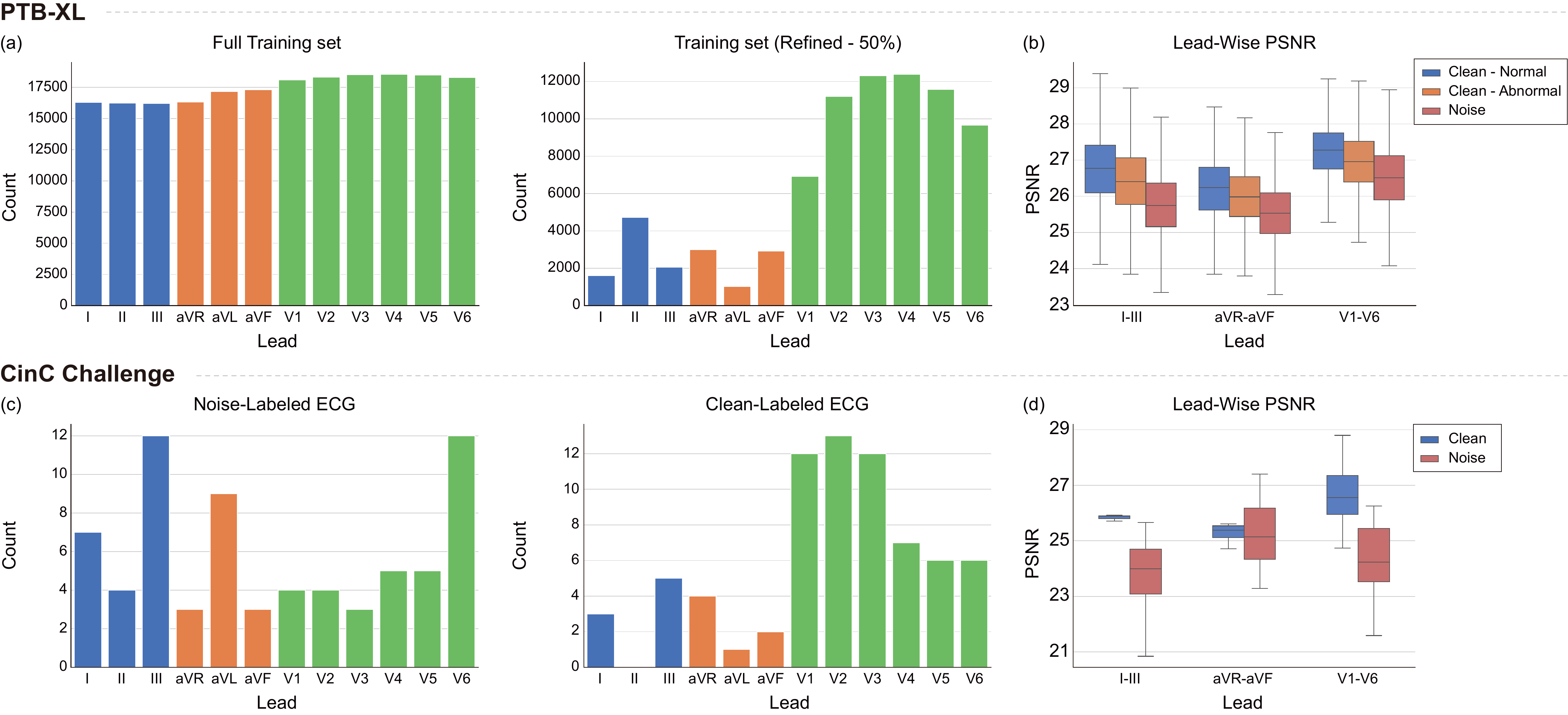}
\caption{
Lead-wise and clinical stratification analysis of ECG signal quality. 
(a) Comparison of lead distributions between the full PTB-XL training set and the top 50\% peak signal-to-noise ratio (PSNR)-ranked subset among clean-labeled recordings. 
(b) PSNR distributions for precordial, limb, and augmented limb leads in PTB-XL, stratified by clean-normal, clean-abnormal, and noise-labeled categories. 
(c) Comparison of lead-type prevalence between clean- and noise-labeled ECGs in the PhysioNet/Computing in Cardiology (CinC) Challenge 2011 dataset (using re-annotated lead-level labels from~\cite{ref10}). 
(d) Average PSNR values by lead group in CinC, based on the same re-annotated labels.
}
\label{fig:figure7}
\end{figure*}

% figure-8
\begin{figure*}[!ht]
\centering
\includegraphics[width=\linewidth]{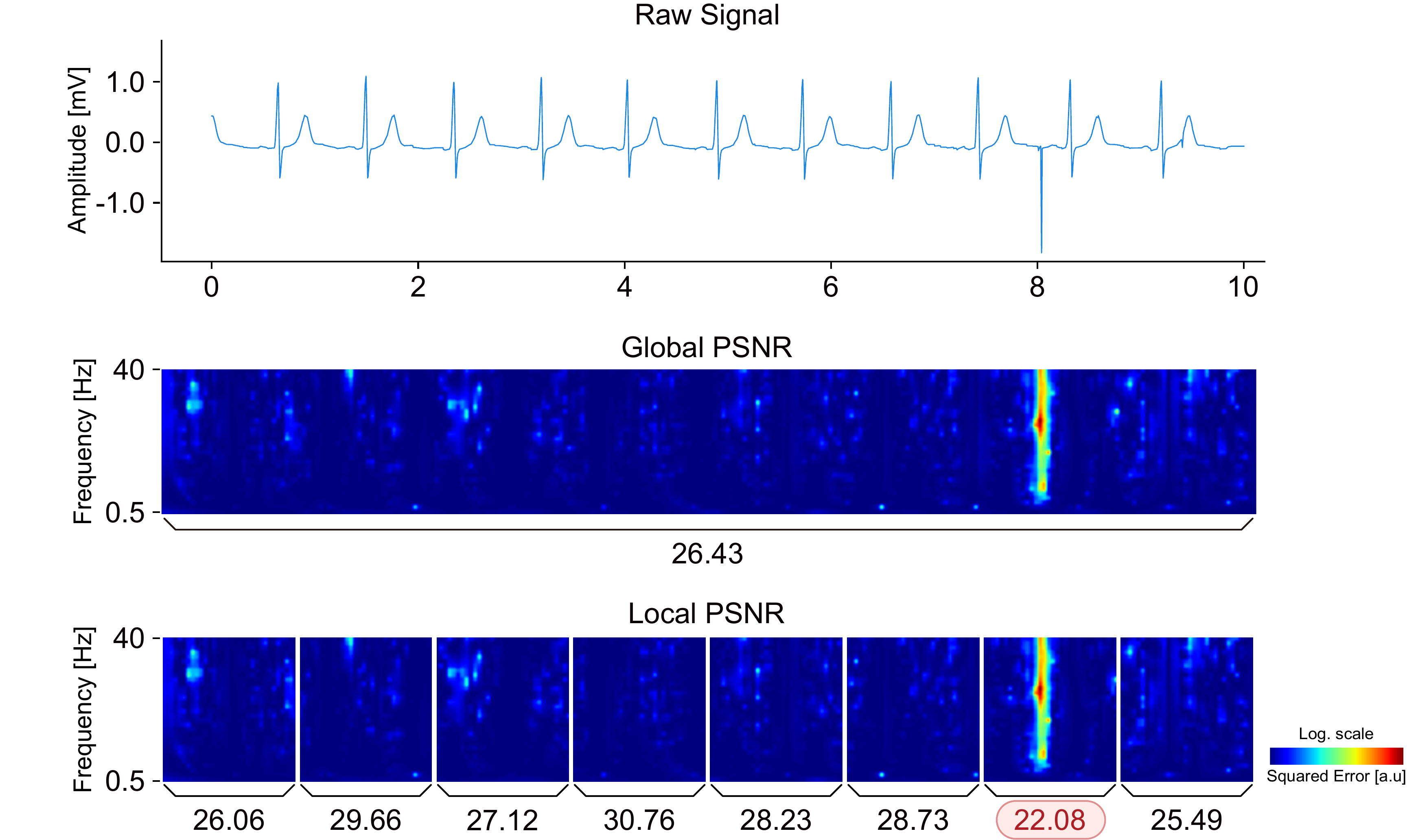}
\caption{
Comparison of global and localized peak signal-to-noise ratio (PSNR) in a 10-s ECG segment. While the global PSNR was 26.43, the localized PSNR in the noise-affected region dropped to 22.08. This demonstrates the utility of sliding-window analysis for detecting short-duration or localized noise that may be overlooked by segment-level averaging.
}
\label{fig:figure8}
\end{figure*}

\subsection*{Clinical Implications}
The proposed method provides substantial clinical benefits by objectively quantifying ECG noise and enhancing diagnostic accuracy. As shown in Figure~\ref{fig:figure6}(a), lower PSNR values are directly correlated with decreased arrhythmia classification accuracy. Clinicians can leverage the continuous quality metric (PSNR) to automatically filter unreliable segments, significantly improving diagnostic workflows, especially in artifact-prone environments such as emergency care and continuous ambulatory monitoring~\cite{ref26}. Furthermore, the effectiveness of our framework in long-term monitoring scenarios, as illustrated in Figure~\ref{fig:figure6}(b), underscores its practical value in sleep studies and chronic patient monitoring, where manual assessment is often infeasible~\cite{ref45}. The early and accurate identification of signal deterioration facilitated by our model could considerably improve patient outcomes by enabling timely clinical interventions.

False alarms remain a significant challenge in intensive care units and telemetry settings, where continuous physiological signal monitoring frequently triggers alerts due to noise rather than genuine clinical events~\cite{ref58,ref59}. By reliably quantifying the signal quality, the proposed method can effectively differentiate between actual physiological deterioration and signal degradation, enhancing diagnostic accuracy, patient safety, and clinical workflow efficiency. Integrating PSNR-based quality scores into existing alarm logic can suppress false positives during periods of high signal noise, thereby reducing unnecessary clinician burden without compromising the detection of clinically meaningful events. In addition, PSNR metrics can generate targeted alerts specifically for nursing staff when signal degradation is detected, prompting appropriate actions, such as lead adjustment or repositioning, rather than emergency interventions. Collectively, these features underscore the potential of the proposed framework to significantly improve the safety, usability, and diagnostic reliability of real-world monitoring systems.

\subsection*{Limitations and Suggestions}
Empirical analyses suggest that a PSNR threshold of approximately 24 reliably distinguishes severely degraded ECG segments. Nevertheless, practical implementation may require adaptive PSNR thresholds tailored to lead-specific characteristics, given the consistently higher signal quality in precordial leads compared with limb leads, as documented in previous studies~\cite{ref33} and corroborated by Figure~\ref{fig:figure7}. In addition, although the global PSNR effectively identifies general noise trends, implementing a sliding-window approach (Figure~\ref{fig:figure8}) could further improve the detection sensitivity for transient and localized noise artifacts, ensuring comprehensive ECG quality assessment in real-time clinical settings. Although these practical modifications offer potential improvements, several limitations remain that warrant further investigation.

Although the final model, utilizing LDM and DDIM, achieves optimal performance within only three reverse diffusion steps, it is still slightly more computationally demanding than traditional methodologies. Furthermore, despite strong generalization across datasets, variability in ECG morphology associated with diverse cardiac pathologies may influence the noise quantification accuracy. In addition, refining the training datasets using PSNR thresholds, although beneficial, risks excluding clinically relevant ECG signals indicative of pathological conditions, which inherently exhibit a lower PSNR despite being noise-free (Figure~\ref{fig:figure7}(b)). Excessive refinement, as illustrated in Figure~\ref{fig:figure4}, could inadvertently bias the model toward healthier populations, thus potentially impairing its effectiveness. Therefore future research should investigate adaptive, clinically informed thresholding strategies and incorporate human-in-the-loop methodologies~\cite{ref63} to balance data quality refinement and accurately represent clinical variability. In addition, further validation using larger and more diverse clinical datasets is recommended to strengthen the robustness and applicability of our proposed framework.

\subsection*{Conclusion}
This study proposed a diffusion model-based framework for ECG noise quantification, effectively addressing critical limitations of traditional and contemporary methods. Our final model significantly outperformed existing SQIs and deep learning approaches, delivering robust and generalizable noise quantification validated across multiple external datasets. By providing an objective, scalable, and efficient solution, our method enhances clinical decision-making, diagnostic accuracy, and real-time ECG monitoring capabilities, thus supporting future advancements in clinical and wearable ECG applications.

\section*{Code availability}
Our codes are available at GitHub: \url{https://github.com/Taeseong-Han/ECGNoiseQuantification}

\section*{Acknowledgements}
This work was supported by the National Research Foundation of Korea (NRF) grant funded by the Korea government (MSIT) (No. 2022R1A2C1013205 and No. RS-2024-00336744).

\section*{Author contributions}
T.S.H. conceptualized the study, developed the methodology, performed data analysis and interpretation, and drafted the manuscript. J.W.H. contributed to data preprocessing, algorithm implementation, and figure preparation. H.K. and C.H.L. assisted with data interpretation and manuscript revision. H.H. and E.K.C. provided clinical expertise and reviewed the results for clinical relevance. D.J.K. supervised the entire project, guided the study design, interpreted the results, and critically revised the manuscript. All authors reviewed and approved the final version of the manuscript.

\section*{Competing interests}
The authors declare no competing interests.

%%%%%%%%%%%%%%%%%%%%%%%%%%%%%%%%%%%%%%%%%%%%%%%%%%%%%%%%%%%%%%%%%%%%%%%%%%%%%%

\bibliography{reference}

\end{document}